\documentclass[useAMS,usenatbib]{mn2e}

\usepackage{graphicx}

\title[Tremaine-Weinberg integrals for double bars]
{Response of the integrals in the Tremaine-Weinberg method to multiple
pattern speeds: a counter-rotating inner bar in NGC 2950?}
\author[W. Maciejewski]
{Witold Maciejewski\thanks{E-mail:witold@astro.ox.ac.uk}\\
Astrophysics, Denys Wilkinson Building, Keble Road, Oxford OX1 3RH}

\begin{document}

\maketitle

\begin{abstract}
When integrals in the standard Tremaine-Weinberg method are evaluated for 
the case of a realistic model of a doubly barred galaxy, their modifications 
introduced by the second rotating pattern are in accord with what can be 
derived from a simple extension of that method, based on separation of 
tracer's density. This extension yields a qualitative argument that 
discriminates between prograde and retrograde inner bars. However, the estimate
of the value of inner bar's pattern speed requires further assumptions. 
When this extension of the Tremaine-Weinberg method is applied to the recent 
observation of the doubly barred galaxy NGC 2950, it indicates that the inner 
bar there is counter-rotating, possibly with the pattern speed of 
$-140 \pm 50$ km s$^{-1}$ arcsec$^{-1}$. The occurrence of counter-rotating 
inner bars can constrain theories of galaxy formation.
\end{abstract}


\section{Introduction}
Bars within bars appear to be a common phenomenon in galaxies. Recent surveys 
indicate that up to 30\% of early-type barred galaxies contain such double bars
(Erwin \& Sparke 2002; Laine et al. 2002). Inner bars remain distinct in near 
infrared (Wozniak et al. 1995), therefore fairly old stars must 
contribute to their light. The relative orientation of the two
bars in doubly barred galaxies is random, therefore it is likely that the 
bars rotate with different pattern speeds. 

The origin of multiply barred systems remains unclear. A bar takes away 
angular momentum from gas very efficiently, and in the young Universe 
bars might have been responsible for the early rapid growth of the massive
black holes
(Begelman, Volenteri \& Rees 2006). If the innermost parts of galactic 
discs formed first, then early instabilities there might have lead to the 
formation of small-scale bars, which may be surviving in the present-day 
Universe as nuclear bars, often nested inside larger, outer bars, that might 
have formed later. Studying the dynamics of nested bars can therefore help us
to understand the formation of galaxies and of their central massive black
holes.

Our understanding of how double bars are sustained has significantly improved
in the recent years. Maciejewski \& Sparke (1997, 2000) have developed a 
formalism that enables finding families of stable regular orbits in such 
systems. Stable regular orbits are robust structures, which define the 
shape of the galaxy, and therefore they can serve as a backbone for double 
bars. So far they have been analyzed for the case when the pattern speed 
of the inner bar is higher than that of the outer bar (Maciejewski \& Sparke 
2000). Using N-body simulations, Rautiainen et al. (2002) confirmed that stars 
get trapped around these orbits, and form long-lasting doubly barred systems. 
However, in this scenario the kinematics of the inner bar is not a scaled-down 
copy of that of the main bar, since the inner bar cannot extend to its 
corotation. Other numerical simulations have shown that systems of two 
counter-rotating bars are also possible (Sellwood \& Merritt 1994;
Friedli 1996), and systems with secondary bars rotating slower than the outer,
main bars, have never been excluded on theoretical grounds. 

Various dynamical scenarios for doubly barred galaxies may have different 
implications for the evolution of the galactic centres. In order to
discriminate between them, one should measure the 
pattern speed of the inner bar. If only one pattern speed is present in the
system, the Tremaine-Weinberg (1984) method allows to derive it using a set of 
simple kinematical measurements. Recent kinematical observations of the
doubly barred galaxy NGC~2950 (Corsini, Debattista \& Aguerri 2003; hereafter 
CDA03) are inconsistent with one pattern speed there. The observations are 
suggestive of another pattern speed in the area of the inner bar. CDA03 
attempted to estimate it, but they concluded a wide range of values, 
consistent with a fast-rotating prograde secondary bar as well as with a 
retrograde one. 

In this paper, I show that a simple extension of the Tremaine-Weinberg method 
to multiple pattern speeds, based on the separation of tracer's density,
is sufficient to discriminate between prograde and retrograde inner bars.
Similar extension has been already considered by CDA03, but they did not 
explore its implications. In Section 2, I outline the
extended method, and I show that it immediately gives a qualitative information
about the sign of rotation of the inner bar. In Section 3, I calculate the
integrals in the original Tremaine-Weinberg method for the case of a realistic
doubly barred galaxy, and I show that their deviation from values for a 
single rotating pattern is the same as predicted by the extension to the 
Tremaine-Weinberg method proposed in Section 2. In Section 4, I examine 
one method to measure the pattern speed of the inner bar, and I show that it 
recovers the pattern speed in the model with an acceptable accuracy. 
Limitations 
of the extended method, other attempts to estimate multiple pattern speeds
using the Tremaine-Weinberg formalism, and consequences of counter-rotating 
inner bars for the evolution of galactic centres are discussed in Section 5.

When the extended Tremaine-Weinberg
method proposed here is applied to NGC~2950, it indicates
that the pattern speed of the inner bar there is smaller than the positive
pattern speed of the outer bar, and that it is most likely negative, i.e. the
inner bar is counter-rotating with respect to the outer bar and to the disc 
(see also Maciejewski 2004).

\section{A simple extension of the Tremaine-Weinberg method to multiple pattern
speeds}
The Tremaine-Weinberg method is designed for one pattern speed, which it 
derives from the luminosity centroid and the luminosity-weighted line-of-sight 
velocity of a chosen tracer moving in the galaxy's potential. The method
rests on three assumptions: the disc of the galaxy is flat, it has a
well-defined pattern speed, and the tracer obeys the continuity equation.
Under these assumptions, the surface density of the tracer, $\Sigma(x,y,t)$,
can be written as 
\begin{equation}
\Sigma(x,y,t) =\tilde{\Sigma} (R, \varphi - \Omega_P t)
\end{equation}
where $\Omega_P$ is the pattern speed, $(x,y)$ are Cartesian coordinates in 
the disc plane, $(R,\varphi)$ are polar coordinates there, centred on the 
galactic centre, and $t$ is time.

When another pattern speed is introduced, the two patterns cannot rotate 
rigidly one through another (Louis \& Gerhardt 1988; Sridhar 1989; Maciejewski 
\& Sparke 2000), and in principle one cannot split the right-hand side
of (1) into components with different pattern speeds. However, if a rough 
separation of patterns is possible, and if one neglects secular evolution,
this system is periodic with period $P=\pi/(\Omega_S - \Omega_B)$, where
$\Omega_S$ and  $\Omega_B$ are pattern speeds of the two bars. Consequently,
the surface density of the tracer can generally be written as
\begin{equation}
\Sigma(x,y,t) =\tilde{\Sigma}_B (R, \varphi - \Omega_B t, t|P) + 
               \tilde{\Sigma}_S (R, \varphi - \Omega_S t, t|P) ,
\end{equation}
where $t|P$ denotes dependence on time with periodicity $P$. In Appendix A,
I evaluate correction to the Tremaine-Weinberg integrals that arises because 
of this periodic oscillation of realistic double bars. This correction turns 
out to be small, and I will neglect it in the following argument.

Once periodic oscillations of nested bars are neglected, (2) gets reduced to
\begin{equation}
\Sigma(x,y,t) =\tilde{\Sigma}_B (R, \varphi - \Omega_B t) + 
               \tilde{\Sigma}_S (R, \varphi - \Omega_S t) .
\end{equation}
Twofold interpretation of such a separation of tracer's density is possible.
Either the tracer in the disc (e.g. stars) can be divided into subgroups 
belonging to two unchanging patterns, each rotating with a constant pattern 
speed, or one can separate radial zones in the galactic disc that rotate with 
constant pattern speeds, with no net exchange of the tracer between the zones. 
Neither of these interpretations is consistent with the dynamics of galaxies,
because patterns rotating one through another change with time, and tracers 
of each pattern are often present at the same radius. However, as I show in 
Appendix A, these inconsistencies are likely to be small, and the separation 
(3) can still be approximately valid.

Below I will show that although the extension of the 
Tremaine-Weinberg method based on the separation of tracer's density (3)
cannot recover the two pattern speeds, it can yield 
a qualitative prediction of how the integrals in the standard 
Tremaine-Weinberg method change when a second rotating pattern is introduced.
In Section 3, I will show that this change, calculated properly for a realistic
model of doubly barred galaxy, with bars oscillating in time, and with
tracers of each bar overlapping, is the same as predicted by 
the extended method proposed here. Thus the separation of tracer's density 
may help in interpreting the Tremaine-Weinberg integrals in the models and 
in the observed galaxies.

Once tracer's density is separated according to (3), one can write two separate
continuity equations. Further derivation is identical to the one performed 
by Tremaine \& Weinberg (1984), and it can be conducted for each tracer's 
density separately, leading to two equations:
\begin{eqnarray}
\lefteqn{\Omega_B \sin i \int_{-\infty}^{+\infty} \Sigma_B (X,Y_{slit}) X dX =}  \nonumber \\
& & \int_{-\infty}^{+\infty} \Sigma_B (X,Y_{slit}) V^{los}_B (X,Y_{slit}) dX , 
\end{eqnarray}
\begin{eqnarray}
\lefteqn{\Omega_S \sin i \int_{-\infty}^{+\infty} \Sigma_S (X,Y_{slit}) X dX =}  \nonumber \\
& & \int_{-\infty}^{+\infty} \Sigma_S (X,Y_{slit}) V^{los}_S (X,Y_{slit}) dX ,
\end{eqnarray}
where $(X,Y_{slit})$ are coordinates on the sky ($X$ running parallel to the 
line of nodes and $Y_{slit}$ being the offset, perpendicular to the line of 
nodes, of the slit along which the integration is performed), $V^{los}$ is the 
observed line-of-sight velocity, and $i$ 
is the inclination of the disc. Let the tracer be made of 
stellar-light emission; then one can define for each bar component, as well 
as for the whole galaxy, the luminosity-density
\[
L_{\diamond} = \int_{-\infty}^{+\infty} \Sigma_{\diamond} (X,Y_{slit}) dX ,
\]
the luminosity centroid
\[
X_{\diamond} = \frac{1}{L_{\diamond}} \int_{-\infty}^{+\infty} \Sigma_{\diamond} (X,Y_{slit}) X dX ,
\]
and the luminosity-weighted line-of-sight velocity
\[
V_{\diamond} = \frac{1}{L_{\diamond}} \int_{-\infty}^{+\infty} \Sigma_{\diamond} (X,Y_{slit}) V^{los}_{\diamond} (X,Y_{slit}) dX ,
\]
where the index ${\diamond}$ is $B$ for the outer bar, $S$ for the inner bar,
and $tot$ for the whole galaxy. Obviously, only values with index $tot$
can be observed, for example by placing a slit parallel to the line of nodes 
(Fig.1), and deriving along it the values of luminosity, centroid and 
line-of-sight velocity, where all integrals include integration over 
the width of the slit. With the definitions above, the sum of (4) and (5) 
takes a form
\begin{equation}
\Omega_B \sin i \, F_B X_B + \Omega_S \sin i \, F_S X_S = F_B V_B + F_S V_S ,
\label{twsum}
\end{equation}
where $F_{\diamond}=L_{\diamond}/L_{tot}$. CDA03 gave a similar equation, but in their
convention there are no fractions $F_{\diamond}$. Note that 
equation (6) is not sufficient to derive two pattern 
speeds $\Omega_{B,S}$ by measuring the luminosity centroid $X_{tot}$ and the 
luminosity-weighted line-of-sight velocity $V_{tot}$ along the slits. Although
$F_B V_B + F_S V_S \equiv V_{tot}$, and it can be measured directly, the 
pattern speeds are combined with the unknown $F_{B,S}$ and $X_{B,S}$.

\begin{figure}
\vspace{-23mm}
\includegraphics[width=\linewidth]{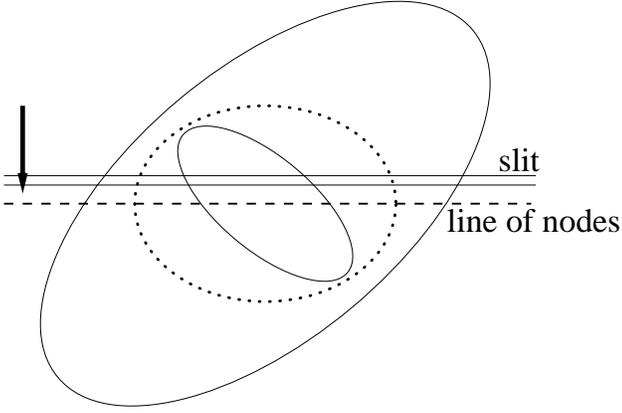}
\vspace{-28mm}
\caption{Schematic drawing of the positions of the bars observed in NGC 2950
and of an exemplary location of the slit. Two solid ellipses outline the
bars, and the dotted ellipse marks the boundary used in the derivation in 
Section 4.}
\end{figure}

However, the form of equation (\ref{twsum}), together with the morphology 
of the doubly barred galaxy, whose pattern speeds we want to measure, already 
reveals some information about the relation between the two pattern speeds. 
The line of nodes passing through the centre of the galaxy sets a division 
of this galaxy into four quadrants. Consider for example the case when the two 
bars lie in opposite quadrants of the galaxy (Fig.1). This is the case
of NGC 2950. Even if we cannot measure 
$X_B$ or $X_S$, we know that they are always of opposite signs, no matter 
how the slit is placed (parallel to the line of nodes). Then if $\Omega_B$ 
and $\Omega_S$ are of the same sign, adding the inner bar should bring the 
sum on the left of (\ref{twsum}) closer to zero (or even through zero for 
strong fast 
inner bars) when compared to the contribution of the outer bar alone. This 
implies that the observed $|V_{tot}|$ should be smaller when entering the 
region of the inner bar than that interpolated from the outer bar. To the
contrary, CDA03 observe $|V_{tot}|$ {\it increasing} around this transition
region in NGC~2950, consistent with $\Omega_S$ having the sign opposite to 
$\Omega_B$, i.e. with the counter-rotating secondary bar.

\begin{figure*}
\includegraphics[width=.48\linewidth]{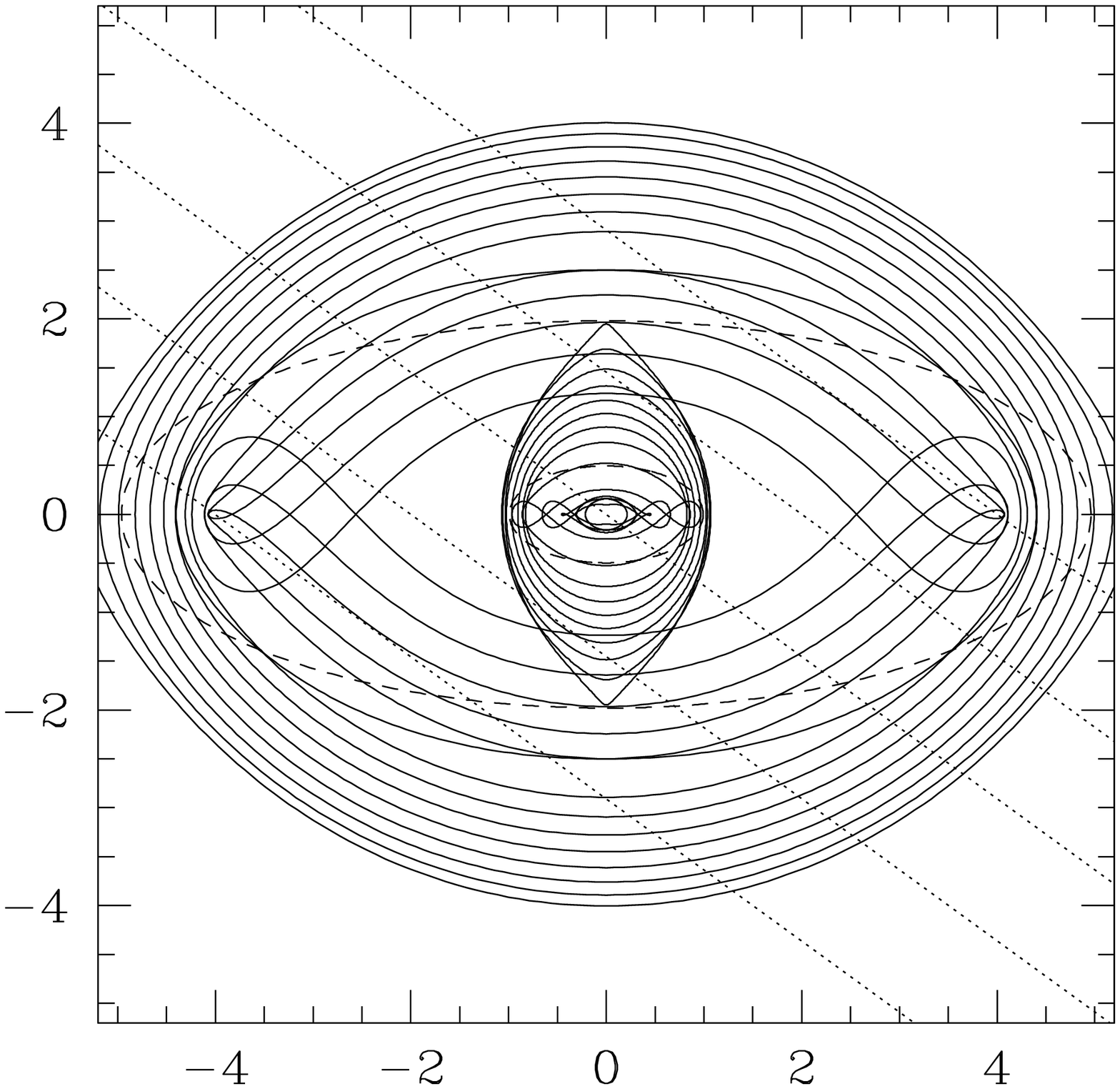}
\includegraphics[width=.48\linewidth]{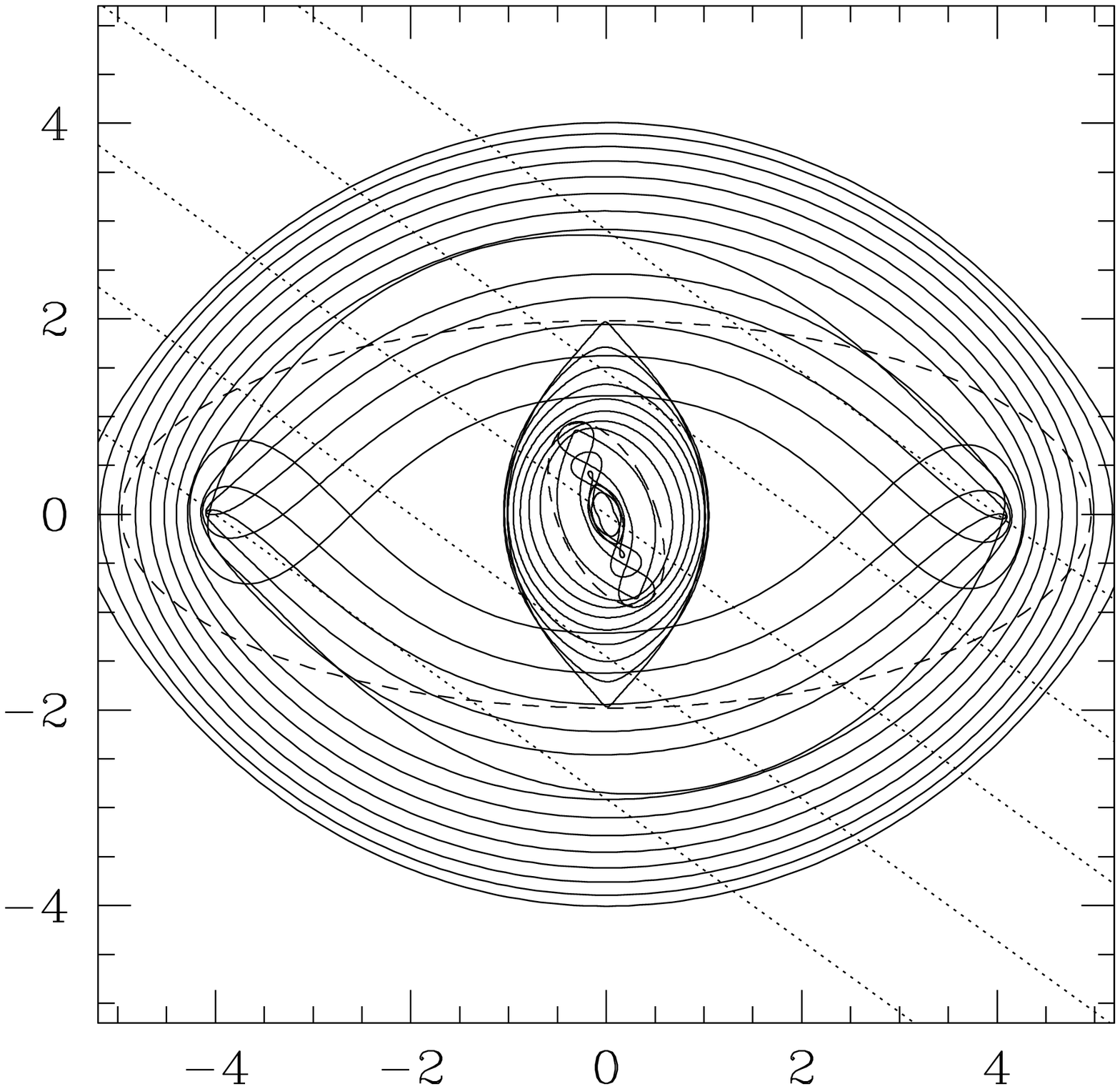}
\vspace{-4mm}
\caption{Loops (solid lines) in a realistic model of doubly barred galaxy 
(Model 2 of Maciejewski \& Sparke 2000) in the linear approximation. Dashed 
ellipses outline the bars: the density of each bar drops there to 0.1 of its 
central value. Dotted lines mark
the direction along which 'slits' are placed, for which the integrated values
of $X_{tot}$ and $V_{tot}$ are derived. The bars are aligned in the left panel,
and at $5\pi/8$ rad angle in the right panel. Units on axes are in kpc.}
\end{figure*}

\subsection{Immediate condition for the sign of $\Omega_S - \Omega_B$}
In the argument above, no reference point has been fixed, since the sums on 
each side of (\ref{twsum}) get modified by the introduction of the secondary 
bar. We do not know the contribution from the outer bar in the region where the
inner bar is present, and interpolation can be misleading. Here I show that 
(\ref{twsum}) allows to tell whether $\Omega_S > \Omega_B$ or 
$\Omega_S < \Omega_B$. Equation (\ref{twsum}) can be rewritten as
\begin{equation}
\Omega_B X_{tot} + X_S F_S (\Omega_S - \Omega_B) = V_{tot}/\sin i .
\end{equation}
This form shows how the relation between the integrals $V_{tot}$ and 
$X_{tot}$ gets modified by the presence of the secondary bar: the second term 
in the sum is a correction term. For slits outside the inner bar this 
correction is zero (because $\Sigma_S$ is zero there), and for those slits 
the observed values of $X_{tot}$ and $V_{tot}$ should lie on a straight line 
of inclination $\Omega_B \sin i $. When the slit passes through the inner bar, 
$X_S \neq 0$. If for such slits $|V_{tot}|$ is larger than the values given 
by the linear relation (7) without the correction factor, then the correction 
has to have the same sign as $\Omega_B X_{tot}$. For the bars that lie in 
opposite quadrants, like in NGC 2950 (Fig.1), 
$X_B$ and $X_S$ are of opposite signs, and $X_{tot}$ has the same sign as 
$X_B$ throughout the galaxy (fig.3 in CDA03). Therefore the two components of 
the sum in (7) can only be of the same sign when $\Omega_B$ has opposite sign 
to $\Omega_S - \Omega_B$. If we take a convention that $\Omega_B>0$ then 
$\Omega_S < \Omega_B$.

This argument can be extended to the case when 
$X_{tot}$ changes sign for slits passing through the secondary bar. In NGC
2950, the values of $X_{tot}$ for these slits are very small, which is
consistent with the observed geometry of bars. A similar argument can be 
applied to galaxies with bars in the same quadrants, where $X_{tot}$ for 
innermost slits is not that small. In this case, $X_B$, $X_S$ and $X_{tot}$
are all of the same sign. If an increase of $|V_{tot}|$ is observed in the
region of the inner bar, $\Omega_B$ must now have the same sign as
$\Omega_S - \Omega_B$, which means $\Omega_S > \Omega_B$ for the assumed 
$\Omega_B>0$. Thus we see that the same increase of $|V_{tot}|$ in the
region of the inner bar can either indicate the inner bar rotating slower or 
faster than the outer bar, depending on the relative orientation of the bars.

\subsection{Condition for counter-rotation}
In some cases, an argument can be made about corotation or 
counter-rotation of the inner bar with respect to the inertial frame.
It relies on the variation of the integrals in the Tremaine-Weinberg method
with the distance of the slit from the centre of the galaxy, $Y_{slit}$. This 
argument can be made under an assumption that $|X_B F_B|$ does not increase 
when we march with the slit through the galaxy toward the line of nodes (in the
direction marked by an arrow on the left of Fig.1). This is a justified 
assumption, since $F_B$ decreases inward for the very reason of the
introduction of the secondary bar, and $|X_B|$ should not go up inward,
since early-type galaxies like NGC 2950 have 'flat' bars with a nearly
constant surface brightness as a function of radius (Elmegreen et al. 1996).

Consider again equation (\ref{twsum}). Normally, for slits that avoid the 
secondary bar, $|V_{tot}|$ decreases inward (i.e. when shifting the slit in the
direction given by the arrow in Fig.1). This is also the case in NGC 2950 
(fig.3 in CDA03). If then $|V_{tot}|$ {\it increases} inward when the slit 
reaches the secondary bar (like in NGC 2950), this can be caused by either of 
the two components of the sum on the left of (\ref{twsum}). I already argued 
that the first component cannot be the cause, since $F_B$ decreases, $|X_B|$ 
is unlikely to increase, and $\Omega_B=const$. The second component can only 
cause the increase of the sum when it is of the 
same sign as the first one. But again, $X_B$ and $X_S$ are of opposite sign. 
Therefore $\Omega_S$ and $\Omega_B$ have to be of opposite sign, too. 

One can repeat this reasoning for a galaxy with bars in the same quadrants,
to show that in that case $|V_{tot}|$ increasing in the region of the inner 
bar indicates that the inner bar is corotating in the inertial frame. Similar
to Section 2.1, the conclusion about the sense of rotation of the 
inner bar depends on the relative orientation of the bars.

\section{Integrals in the Tremaine-Weinberg method calculated for a realistic
model of a doubly barred galaxy}
Regular motion of a particle in a potential of a doubly barred galaxy has
two frequencies associated with it, each related to one of the bars,
in addition to the frequency of its free oscillations (Maciejewski \& Sparke 
1997). In the linear approximation this motion corresponds to epicyclic 
oscillations with these frequencies around the guiding radius. Particles 
with the same guiding 
radii are bound to closed curves (loops) that oscillate in the pulsating
potential of double bars. Loops are also observed in nonlinear
analysis (Maciejewski \& Sparke 2000; Maciejewski \& Athanassoula 2006).

The orbital approach directly indicates that the motion 
of a particle associated with one bar has always a component coming from
the other bar. Amplitudes of the oscillations can be easily evaluated in the
linear approximation (Maciejewski 2003), giving particle's position 
and velocity at each relative position of the bars. In Fig.2, I plot the loops
populated by particles moving in the potential of a doubly barred galaxy 
constructed by Maciejewski \& Sparke (2000; Model 2). This is a realistic 
potential,
since it admits orbits that support the outer bar, as well as orbits supporting
the inner bar, throughout the extent of each bar. The two bars in this model
rotate in the same direction with  pattern speeds $\Omega_B = 35$ km s$^{-1}$ 
kpc$^{-1}$ and $\Omega_S = 110$ km s$^{-1}$ kpc$^{-1}$. I plot the loops for 
two relative orientations of the bars: the bars parallel (Fig.2, left), and the
bars at the angle of $5\pi/8$ rad, the value similar to that observed
in NGC~2950 after deprojection (Fig.2, right). The loops change shapes as the 
bars rotate one with respect to another, and loops associated with one bar 
intersect the loops associated with the other bar. Thus formally one cannot
perform here the separation of tracer's density postulated in (3).

\begin{figure*}
\includegraphics[width=.48\linewidth]{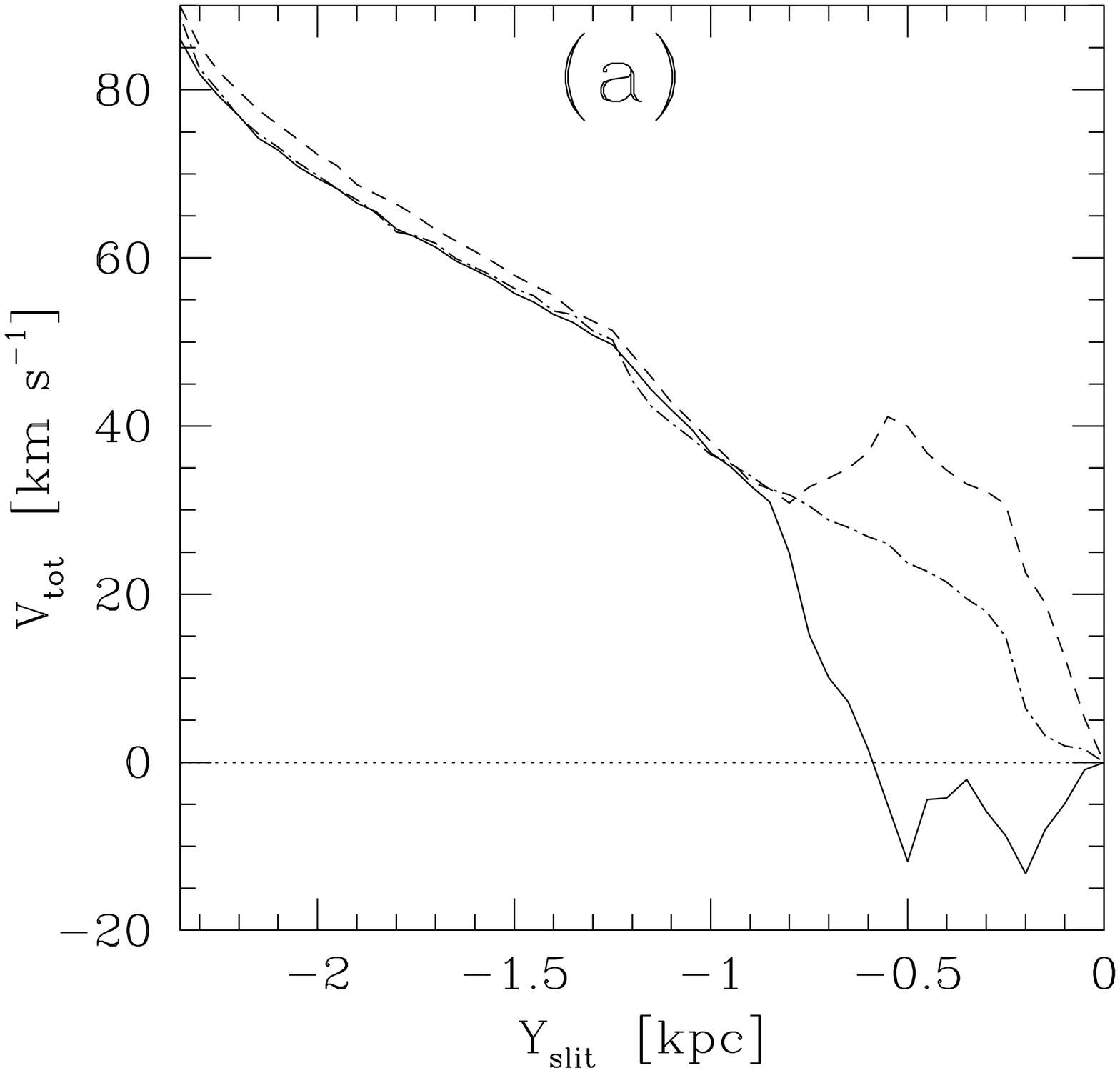}
\includegraphics[width=.48\linewidth]{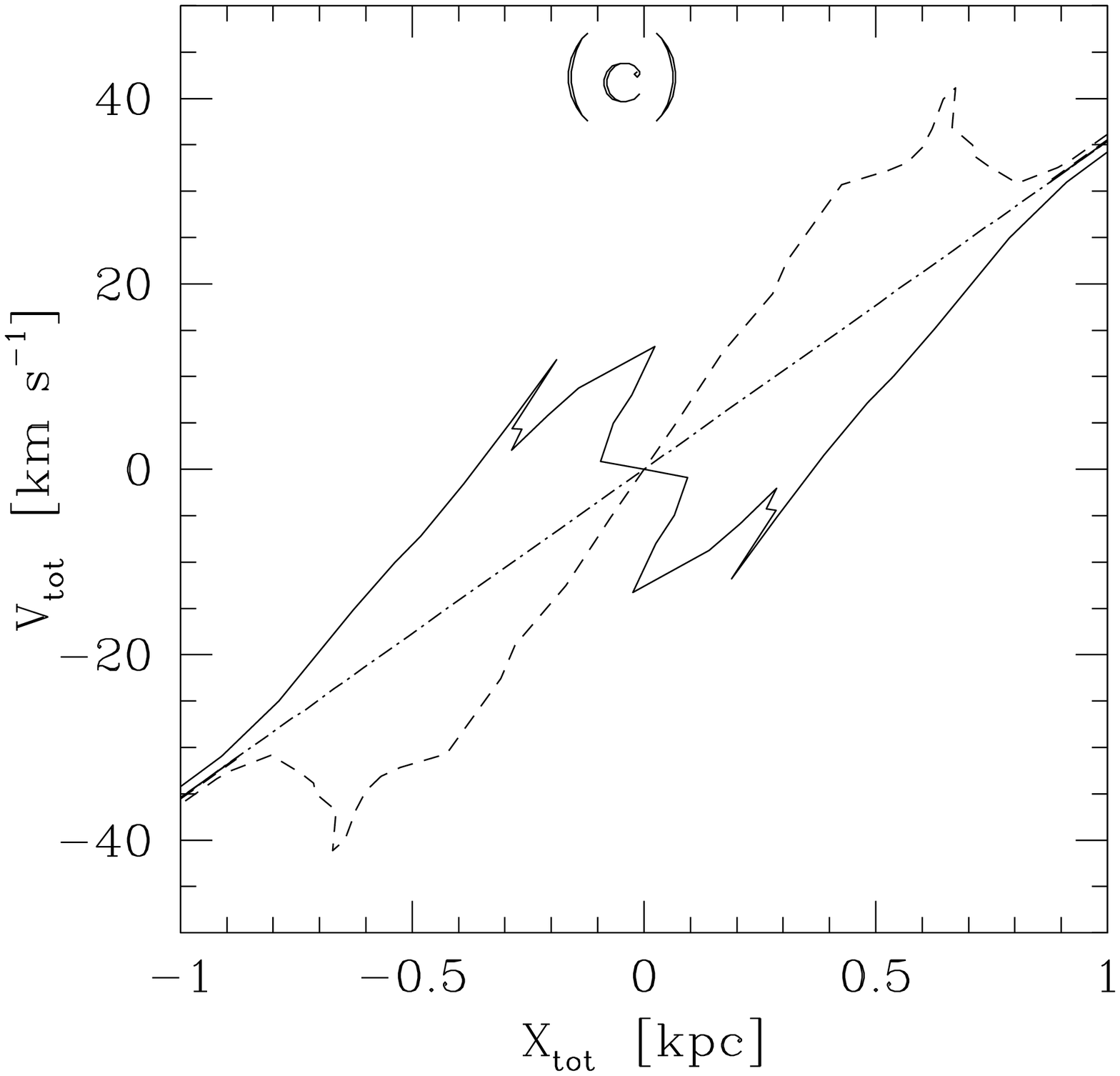}
\includegraphics[width=.48\linewidth]{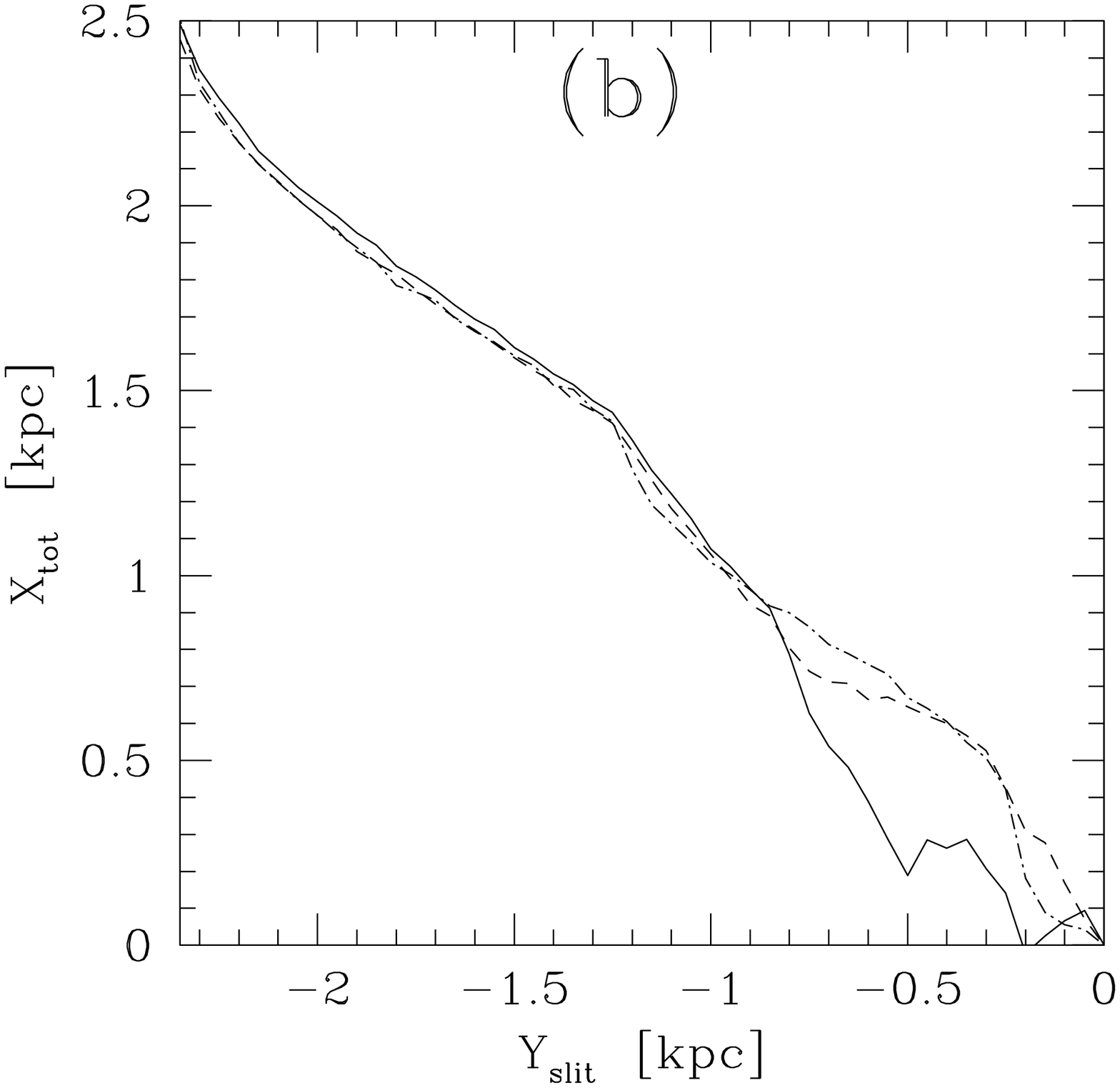}
\includegraphics[width=.48\linewidth]{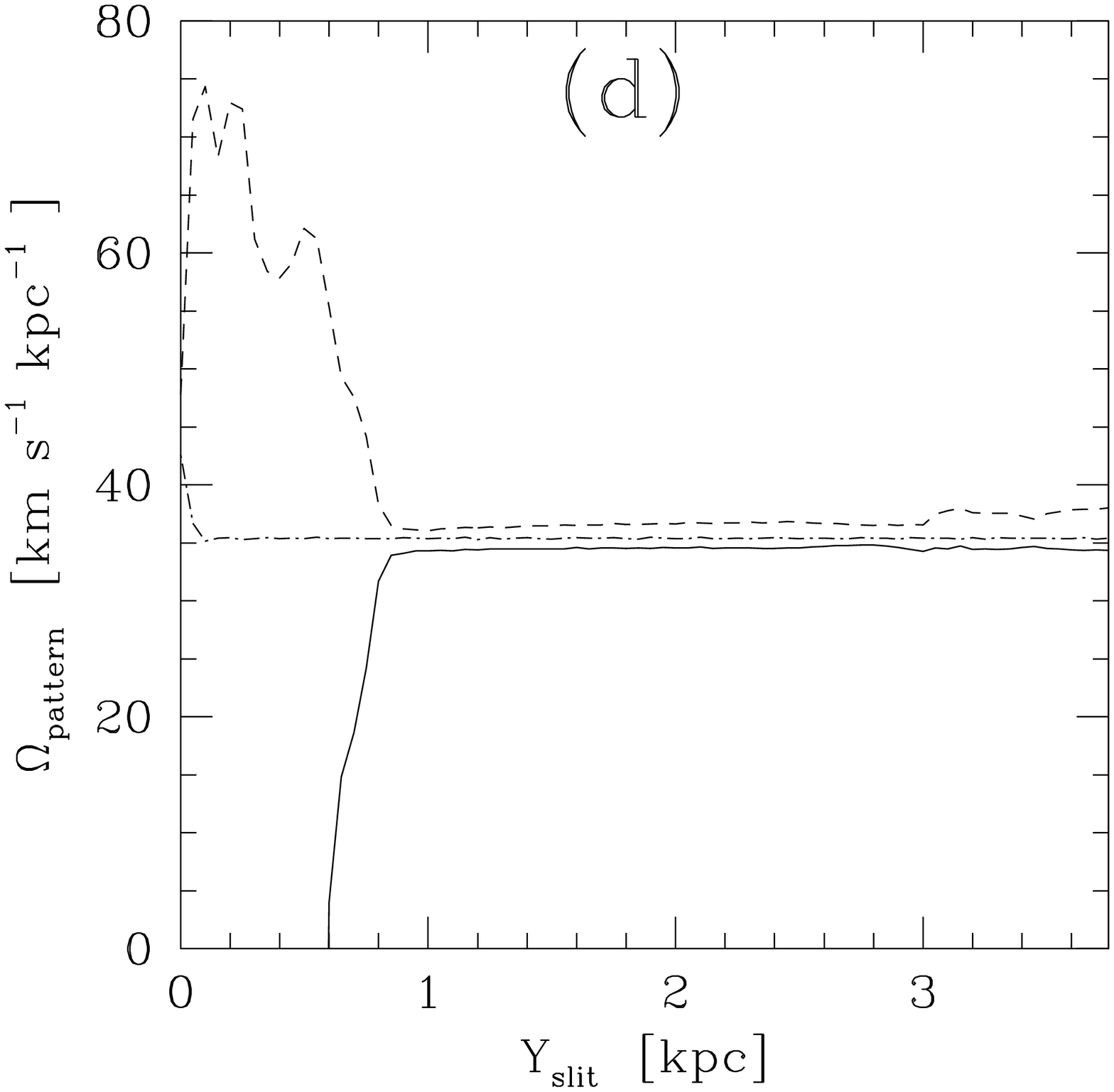}
%\vspace{-4mm}
\caption{Values of the integrals in the Tremaine-Weinberg method, calculated 
from the epicyclic formulae for the positions and velocities of particles in 
the realistic model of doubly barred galaxy from Fig.2. The dot-dashed line 
corresponds to the model without the inner bar, the dashed line is for the
model with two bars aligned, as in the left panel of Fig.2, while
the solid line is for the model with two bars located as in the right panel 
of Fig.2. This figure can be directly compared with fig.3 in CDA03.
{\it (a)} The kinematic integrals $V_{tot}$, defined in Section 2, 
as a function of the slit offset $Y_{slit}$ with respect to the centre of the 
galaxy. {\it (b)} The photometric integrals $X_{tot}$, defined in Section
2, as a function of the slit offset $Y_{slit}$. {\it (c)} $V_{tot}$ as a
function of $X_{tot}$. {\it (d)} Pattern speed in the model, 
$\Omega_{pattern} \equiv V_{tot}/X_{tot}$, derived
separately for each position of the slit $Y_{slit}$.}
\end{figure*}

However, changes in the shapes of the bars, and zones where both tracers 
coexist, may be small, and then, what formally prohibits the separation, may 
turn into a higher-order correction to it (see also Appendix A). In order to 
check whether it is the case for the realistic Model 2 (Maciejewski \& Sparke 
2000), I calculated the positions and velocities of some $10^6$ points on 200 
loops in that model, from which I obtained the centroids $X_{tot}$ and the 
line-of-sight velocities $V_{tot}$ along slits placed at the same relative 
angle to the outer bar as in the CDA03 observations of NGC~2950 (dotted 
straight lines in Fig.2). For clarity, only 28 loops are displayed in Fig.2,
out of 200 used in the calculations. The 200 loops were populated with 
particles, so that a smooth density distribution in the bars is recovered.
Several recipes have been adopted to populate loops with particles, in order
to make sure that the results below do not depend on the way in which the model
is constructed. In the model, I use deprojected velocities,
which replaces $V_{tot}/\sin i$ by $V_{tot}$, and I measure $Y_{slit}$ in
the plane of the galaxy.

In Fig.3b, I plot the value 
of the centroid, $X_{tot}$, as a function of the offset of the slit from the 
galaxy centre, $Y_{slit}$. When the inner bar is placed at the position 
similar to the one observed in NGC~2950, the value of $X_{tot}$ (solid line) 
is closer to zero in the region of the inner bar ($|Y_{slit}|<1$), than when 
the inner bar is absent (dot-dashed line). However, $X_{tot}$ does not change 
the sign in the region of the inner bar, which is also observed in NGC~2950 
(fig.3 in CDA03). This indicates that the light integrated along a slit 
passing through the inner bar is still dominated by the outer bar. In the 
case of the two bars parallel (Fig.2, left panel) there is no such decrease 
of $X_{tot}$ in the region of the inner bar (dashed line in Fig.3b).

In Fig.3a, I plot the luminosity-weighted line-of-sight 
velocity, $V_{tot}$, as a function of the same offset of the slit from the 
galaxy centre, $Y_{slit}$, as in Fig.3b. When the inner bar,
rotating in the same direction as the outer bar, but faster, is placed at 
the position similar to the one observed in NGC~2950, the value of $V_{tot}$ 
(solid line) rapidly approaches zero in the region of the inner bar 
($|Y_{slit}|<1$), and in most of this region it has the sign opposite to the
value of $V_{tot}$ for slits not passing through the inner bar 
($|Y_{slit}|>1$). This behaviour of $V_{tot}$ is opposite to that observed
in NGC~2950 by CDA03. For the model of galaxy with $\Omega_S > \Omega_B$,
changes of $V_{tot}$ with $Y_{slit}$ observed by CDA03 can only be reproduced
when both bars lie in the same quadrants of the galaxy. For the particular
case of bars parallel, presented here,
$|V_{tot}|$ increases in the region of the inner bar (dashed line in Fig.3a), 
relative to the case when there is no inner bar (dot-dashed 
line). This dependence of deviations in $V_{tot}$ on the relative position
of the two bars is exactly as expected from the simple
extension of the Tremaine-Weinberg method derived Section 2, based on
the separation of tracer's density (3). Namely, if the
inner bar rotates faster than the outer bar then, when both bars are in the 
same quadrants (defined by the line of nodes and the centre of the galaxy),
$|V_{tot}|$ increases in the slits passing through the inner bar, but it 
{\it decreases}, when the bars are in the opposite quadrants. 

The plot of $V_{tot}$ as a function of $X_{tot}$ is presented in Fig.3c. One 
may attempt to fit a second straight line in the region of the inner bar, but 
this fit will not yield the actual pattern speed of the inner bar. In
Fig.3d, I plot the ratios of $V_{tot}/X_{tot}$ as estimators of the derived
$\Omega_{pattern}$. For a single bar (dot-dashed line), there is one pattern
speed independent of the offset of the slit, and equal to the one assumed
in the model. If there are two pattern speeds in the system,
the inner bar does not alter significantly the pattern speed of the outer 
bar derived from the slits that do not pass through this inner bar. However,
the induced deviations are systematic: pattern speed of the outer bar is
spuriously increased when the bars lie in the same quadrants, and decreased
when they lie in quadrants opposite. The effect is small though, below 5\%.
For the slits passing through the inner bar, the derived $\Omega_{pattern}$ 
rapidly decreases if the bars lie in opposite quadrants (solid line), and it 
becomes negative, not giving any information about the
pattern speed of the inner bar. If the bars lie in the same quadrants (dashed
line), the derived $\Omega_{pattern}$ is larger than $\Omega_B$ for the slits
passing through the inner bar, but it varies with $Y_{slit}$, and never reaches
the value of $\Omega_S=110$ km s$^{-1}$ kpc$^{-1}$ assumed in the model, 
remaining at slightly over half of that value.

\begin{figure}
\includegraphics[width=\linewidth]{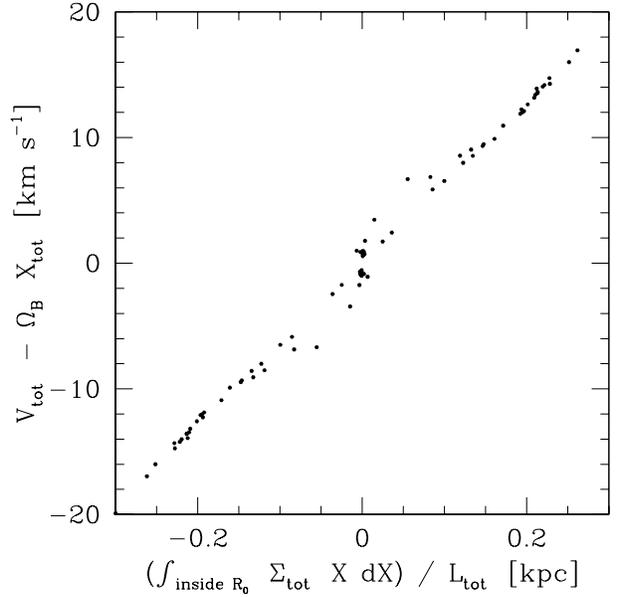}
\caption{Model values of the integrals that determine the linear regression in 
(10) with the 
expected slope of $\Omega_S - \Omega_B$. Values of $\frac{1}{L_{tot}} 
\int_{-X_0}^{X_0} \Sigma_{tot} (X,Y_{slit}) X dX$ are plotted against 
$V_{tot} - \Omega_B X_{tot}$.}
\end{figure}

\section{A method to derive the value of $\Omega_S$}
The main goal of this paper is to show that with a simple extension of 
the Tremaine-Weinberg method to multiple pattern speeds one can derive at 
least rough qualitative information about the secondary pattern rotation.
For the data on NGC~2950 presented by CDA03 this information is that the
inner bar rotates in the opposite direction than the outer bar.
On the other hand, getting the numerical value of the pattern speed of the 
secondary bar, $\Omega_S$, may not be possible without making additional 
assumptions. Here I analyse one possible method to calculate $\Omega_S$.

If the tracer of the secondary bar does not extend beyond some radius $R_0$ 
in the galaxy plane (outlined in projection by the dotted ellipse in Fig.1), 
then (5) can be rewritten as
\begin{eqnarray}
\lefteqn{\Omega_S \sin i \int_{-X_0}^{X_0} \Sigma_S (X,Y_{slit}) X dX =} \nonumber \\
& & \int_{-X_0}^{+X_0} \Sigma_S (X,Y_{slit}) V^{los}_S (X,Y_{slit}) dX ,
\end{eqnarray}
where $R_0^2 = (Y_{slit}/\cos i)^2 + X_0^2$. Summing (4) and (8) leads to
\begin{equation}
\Omega_B X_{tot} + \frac{\Omega_S - \Omega_B}{L_{tot}}
\int_{-X_0}^{X_0} \Sigma_S (X,Y_{slit}) X dX = \frac{V_{tot}}{\sin i}.
\end{equation}
However, aside for $\Omega_S$, we still do not know the integral in (9). It 
can be approximated when one assumes that only tracers of the inner bar and 
of the axisymmetric component are present inside the dotted ellipse in Fig.1, 
i.e. that the outer bar is almost axisymmetric in this region. Note that this 
is a strong and poorly founded assumption. However, if we take it, and since 
the contribution of the axisymmetric component to the integral in (9) cancels
out, one can substitute there the observable $\Sigma_{tot}$ for the unknown 
$\Sigma_S$, and rewrite (9) as a linear regression of $\Omega_S - \Omega_B$:
\begin{equation}
\frac{\Omega_S - \Omega_B}{L_{tot}} 
\int_{-X_0}^{X_0} \Sigma_{tot} (X,Y_{slit}) X dX =
\frac{V_{tot}}{\sin i} - \Omega_B X_{tot} .
\end{equation}
Thus $\Omega_S - \Omega_B$ can be obtained as a slope of a straight line 
fitted to the data from slits passing through the inner bar.
A similar equation has been used by CDA03
to estimate $\Omega_S$, but here the coefficient at $\Omega_S - \Omega_B$
is defined differently.

I tested equation (10) on the model examined in Section 3, for the position
angles of the bars like the ones observed in NGC~2950. As in Section 3, 
in the model I use deprojected velocities, which replaces $V_{tot}/\sin i$ 
by $V_{tot}$, and I measure $Y_{slit}$ in the plane of the galaxy. On the basis
of the shapes of the loops, I chose $R_0=1.1$ kpc for this model, and I fixed 
$\Omega_B=35$ km s$^{-1}$ kpc$^{-1}$. In Fig.4, I plot
the data points for the regression (10). They follow a straight line well,
except for the points around zero values. However, these points come from
$Y_{slit} \simeq R_0$, where the integration is over a small number of
particles, hence random error is large there. If we exclude these points,
the derived $\Omega_S - \Omega_B$ oscillates between 60 and 70 km s$^{-1}$ 
kpc$^{-1}$ for $|Y_{slit}|<0.75$ kpc. This gives $\Omega_S$
between 95 and 105 km s$^{-1}$ kpc$^{-1}$, consistent with the input value
of 110 km s$^{-1}$ kpc$^{-1}$, with an error of about 10\%.

In order to apply the same method to the observed data on NGC~2950 from CDA03,
the value of $R_0$ can be estimated by checking how the integral in (10) 
changes with varying $R_0$ --- it should have a plateau in the transition 
area between the bars. I used the R-band image of NGC 2950 (Erwin \& Sparke 
2003) to calculate this integral and $L_{tot}$ for each of the slits passing
through the secondary bar, placed at positions reported by CDA03. The values 
of $\frac{1}{L_{tot}} \int_{-X_0}^{X_0} \Sigma_{tot} (X,Y_{slit}) X dX$ 
as a function of $R_0$ are plotted in Fig.5. All curves indeed have a 
plateau around the same $R_0$ of 5--6 arcsec. It is located just outside 
a nuclear stellar ring at $\sim 4.2$ arcsec, reported by Erwin \& Sparke 
(2003), which is likely circular in the plane of the galaxy. Thus the 
method proposed here may be particularly well suited for NGC 2950.

\begin{figure}
\includegraphics[width=\linewidth]{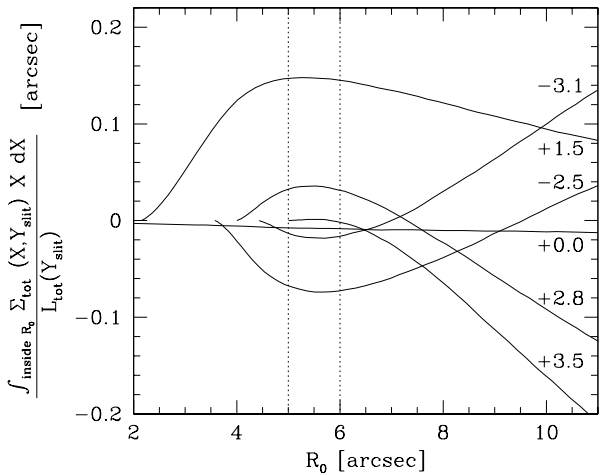}
\vspace{-20mm}
\caption{Values of $\frac{1}{L_{tot}} \int_{-X_0}^{X_0} \Sigma_{tot} 
(X,Y_{slit}) X dX$ as a function of 
$R_0 = \sqrt{(Y_{slit}/\cos i)^2+X_0^2}$ for NGC 2950. The curves are derived 
from the R-band image of NGC 2950 (Erwin \& Sparke 2003), for slits placed 
parallel to the line of 
nodes, and offset from it by values given on the right of the plot in arcsec. 
Two dotted vertical lines indicate a plateau range, common for all the curves.}
\end{figure}

Each curve in Fig.5 has a characteristic U-shape, whose depth depends
on the relative contribution of the inner bar to the integral in (10), and
therefore it vanishes outside the inner bar, and also very close to the line
of nodes, where the integral tends to be zero. From the R-band image, the
depth is largest for slits offset by $\sim \pm 1$ arcsec. The linear fit to 
(10), with the data for $V_{tot}$ and $X_{tot}$ from each slit passing through 
the inner bar provided by CDA03, reveals a large negative value of 
$\Omega_S - \Omega_B$, about $-150 \pm 50$ km s$^{-1}$ arcsec$^{-1}$. 
For $\Omega_B \simeq 11$ km s$^{-1}$ arcsec$^{-1}$, measured by CDA03, this
gives $\Omega_S \simeq -140 \pm 50$ km s$^{-1}$ arcsec$^{-1}$. This result,
although with much poorer grounds than the qualitative arguments 
from Section 2, reaffirms counter-rotation of the inner bar. 
Note that the unrealistically large value of $\Omega_S$ derived with this 
method is likely an overestimate: in order to get it, the integral 
$\int_{-X_0}^{X_0} \Sigma_{tot} (X,Y_{slit}) X dX$ has been substituted for
$\int_{-X_0}^{X_0} \Sigma_S (X,Y_{slit}) X dX$. The substituted integral is 
most likely smaller than the original one, because the contributions of the two
bars are of opposite signs.

\section{Discussion}
The original Tremaine-Weinberg method, and its extension proposed here,
are based on the continuity equation applied to a tracer moving in the
gravitational field of the galaxy. If old stars are taken as the tracer,
the continuity equation is well satisfied globally. If tracers associated 
with each bar are separated as proposed in (2), continuity of each tracer
can only be violated when there is a net flux of stars from one bar to the 
other, consistent over many rotations of the bars. This corresponds to
a secular strengthening of one bar at the cost of the other, hence such a
system is no longer periodic. However, if the mass transfer is slow, the 
analysis presented in this paper is still applicable. This can be 
supported by an argument similar to that presented in Appendix A, but for
densities of the bars monotonically changing. Moreover, studies of orbits
in self-consistent models of double bars show that mass transfer between the
bars is likely to be small, because orbits that are trapped and oscillate 
around one or the other bar, populate large fraction of phase space 
(Maciejewski \& Sparke 2000; Maciejewski \& Athanassoula 2006).

Note that the separation of tracers in (2) is only formal, and (2) applies to
any system that is periodic with period $P$. However, in order to follow the
Tremaine-Weinberg formalism for each tracer, one has to approximate each tracer
as a solid-body rotator, which leads to (3). The analysis of a realistic doubly
barred galactic potential, presented in Section 3, indicates that although
this approximation is incorrect in a rigorous sense, deviations from solid-body
rotation are small, and they only contribute to higher-order terms in the
Tremaine-Weinberg integrals. Namely, even if each particle in the system 
oscillates with frequencies related to both bars, for most of the particles 
one frequency dominates, and the separation can be performed. Although the
presence of two forcing frequencies leads to noticeable pulsation of the inner
bar (e.g. Rautiainen et al. 2002), the magnitude of additional terms in the
Tremaine-Weinberg integrals, that this pulsation gives rise to, is much
smaller than the magnitude of the leading terms, as argued in Appendix A.
The radial separation of the tracers, necessary for the estimate of the 
pattern speed of the inner bar in Section 4, can also be possible, if the 
zones, where two tracers coexist, occupy small fraction of the galaxy. The 
larger this fraction, the poorer is the estimate of the inner bar's pattern 
speed. Further testing of this method will require a fully self-consistent 
model with an inner bar.

This paper shows that the change in value of the integrals $X_{tot}$ and 
$V_{tot}$ in the original Tremaine-Weinberg formalism, caused by the second 
pattern speed, depends on the relative position
of the bars. In particular, I showed that fitting another straight line in 
the $X_{tot} - V_{tot}$ diagram to the data from slits passing through
the inner bar {\it does not} yield the pattern speed of this bar. Therefore
recent derivations of multiple pattern speeds based on such analysis (Rand
\& Wallin 2004; Hernandez et al. 2005) have to be treated with caution.

The retrograde rotation of the inner bar in NGC~2950 proposed here is 
inconsistent with the first estimate of $\Omega_S$ by CDA03. However, in
that estimate it is assumed that the inner bar dominates the light in the slits
passing through it. This cannot be correct, since $X_{tot}$ and $X_S$ are of 
opposite sign there, which indicates that the light from the outer bar still
dominates in the slits passing through the inner bar. The second estimate of 
CDA03, like the estimate given in this paper, indicates that the inner bar 
is counter-rotating.

The obvious next step in verifying counter-rotation of the inner bar in
NGC~2950 would be an examination of orbital structure of a galaxy with a
retrograde inner bar, following the method presented in Section 3. However,
a bar counter-rotating with respect to its stellar disc has to be built out of
so called $x_4$ orbits, which are normally very close to circular (see e.g.
Sellwood \& Wilkinson 1993). In terms of the linear approximation of Section 3,
the amplitude of oscillation around the guiding radius generated by a 
retrograde bar is very small (see equations (12) and (13) in Maciejewski 
2003). Thus a construction of a retrograde bar in a prograde stellar disc
is unlikely. This is consistent with N-body simulations, in which multiple 
pattern speeds form naturally in stellar discs (Rautiainen et al. 2002) --- 
these simulations do not recover retrograde pattern speeds.

This leads to another possibility, namely that the inner bar is
formed out of an inner disc that counter-rotates with respect to the outer 
disc. Numerical results indicate that two counter-rotating bars, formed in
two counter-rotating stellar discs that overlap each other, can survive
in galaxies for many rotation periods (Sellwood \& Merritt 1994; Friedli 
1996). If this is the case, one should find out whether a retrograde
inner disc is consistent with the rotation curve and velocity dispersion
in NGC~2950. The data presented by CDA03 do not indicate counter-rotation in 
the innermost few arc-seconds of NGC~2950. 

If further kinematical studies of NGC~2950 confirm counter-rotation in its 
innermost parts, then extending the Tremaine-Weinberg method along the lines 
proposed here may yield an efficient way of detecting counter-rotation in
galaxies with nested bars. Thus far all detections of counter-rotation in
galaxies come directly from the observed velocity fields obtained with the 
long-slit (e.g. Kuijken, Fisher \& Merrifield 1996) or integral-field 
(e.g. Emsellem et al. 2004) spectroscopy. However, if the 
counter-rotating population is small, it may not be recognized with those 
methods. On the other hand, the Tremaine-Weinberg method goes beyond the raw 
observed velocity field by finding integrals that are useful in detecting a
rotating pattern. In the presence of rotating patterns counter-rotation should 
be spotted more easily with this method.

The use of this method, like of the original Tremaine-Weinberg method, is 
limited to galaxies with bars considerably inclined to both the major and
minor axes of the disc. NGC~2950 fulfils this requirement particularly well,
but there is a number of other galaxies to which this method can be readily 
applied. For example NGC~3368, NGC~3941, NGC~5365, NGC~5850, and NGC~6684 all
have the position angles of both bars separated by 20 to 70 degrees from
the position angle of the line of nodes (see the catalogue by 
Erwin 2004 for the parameters). The bar inside the oval in NGC~3081 has also
parameters favourable for this method. In addition, a modification of the 
Tremaine-Weinberg method along the lines proposed in Section 4 can be applied 
to the interiors of nuclear rings that often host nuclear bars (e.g. NGC~1097, 
NGC~6782).

The fraction of counter-rotating inner bars can
constrain theories of galaxy formation and evolution. Currently the most
accepted view on the origin of the inner bar is that it forms through
instabilities in gas inflowing along the outer bar (Shlosman, Frank \&
Begelman 1989). However, if inner bars form on early stages of galaxy
assembling, and outer bars form from material that settled on the galaxy
later, the spins of the two bars may be unrelated, leading to a large fraction
of counter-rotating inner bars. Determining this fraction may help to 
distinguish between these two evolutionary scenarios.

\section{Conclusions}
In this paper I presented an argument that a simple extension of the 
Tremaine-Weinberg method to multiple pattern speeds can provide an information
about the sense of rotation of the inner bar in doubly barred galaxies.
The extended formula advocated here links the relative position of the 
bars to the sign of rotation of the inner bar. This extension cannot be as 
rigorous as the original method, because it assumes that the patterns do not
change as they rotate one through another, which is not true in general. 
However, it predicts the same deviations of the integrals in the 
Tremaine-Weinberg method from their values for the single rotating pattern, 
as in the orbital model of a realistic doubly barred galaxy, which does not 
involve this assumption. This indicates that the degree of change that the 
patterns rotating one through another undergo is small, and that it does not
affect significantly the results of the extended method proposed here.

Application of the extended method to NGC~2950 implies that the inner bar 
there counter-rotates with respect to the outer bar and to the large-scale 
disc. Since a retrograde bar is unlikely to be supported in a prograde disc,
a retrograde inner disc may be hiding in the central kiloparsecs of NGC~2950.

{\bf Acknowledgements.} I wish to thank Peter Erwin for letting me use the
R-band image of NGC 2950, which he obtained with the WYIN telescope, and for 
the list of other galaxies, to which this method can be applied. I am
grateful to the authors of the CDA03 paper for sharing the details of their
observations with me. Discussions with Linda Sparke improved presentation of 
this argument. This work was partially supported by the Polish Committee for 
Scientific Research as a research project 1 P03D 007 26 in the years 
2004--2007.

\appendix

\section{Correction terms in Tremaine-Weinberg integrals for a pulsating 
Ferrers' bar}

Consider a Ferrers' bar with major and minor axes $a$ and $b$. Its surface 
density can be written in the Cartesian coordinates $(x,y)$ in the plane of 
the galaxy as
\begin{equation}
\Sigma(x,y) = \left\{ \begin{array}{ll}
\Sigma_{0B} \left(1-\frac{x^2}{a^2}-\frac{y^2}{b^2} \right) 
    & \mbox{if $\frac{x^2}{a^2}+\frac{y^2}{b^2} \le 1$} \\
 0  & \mbox{if $\frac{x^2}{a^2}+\frac{y^2}{b^2} > 1$ ,}
\end{array} 
\right.
\end{equation}
where $\Sigma_{0B}$=const is the central density of the bar. Pulsation of the 
bar can be described by periodic variation of the length of its axes
\[ 
a^2(t) = a_0^2 / g_1(t|P) ,  
\qquad b^2(t) = b_0^2 / g_2(t|P) , 
\]
where $g_1$ and $g_2$ are periodic functions of time with period $P$. 
Thus non-zero density, corresponding to the top line in (A1), of a pulsating 
bar that rotates with pattern speed $\Omega_B$, can be written in polar 
coordinates $(R, \varphi)$ as
\begin{eqnarray}
\lefteqn{\Sigma(x,y,t) = \tilde{\Sigma} (R, \varphi - \Omega_B t, t|P) = \Sigma_{0B} \times} \\
& & \left[ 1 - f_1(R, \varphi - \Omega_B t) g_1(t|P)
             - f_2(R, \varphi - \Omega_B t) g_2(t|P) \right] , \nonumber
\end{eqnarray}
where 
\[
f_1(R,\phi) = \frac{r^2 \cos^2\phi}{a_0^2} , 
\qquad f_2(R,\phi) = \frac{r^2 \sin^2\phi}{b_0^2} . 
\]
Note that (A2) is one of the components of (2) for the particular case of a
Ferrers' bar.

Let (A2), wherever larger than zero, describe non-zero surface density of a 
tracer in a rotating and pulsating bar. In the Tremaine-Weinberg method, the
time derivative of this surface density enters the continuity equation. 
Simple partial derivation of (A2) gives
\begin{equation}
\frac{\partial \Sigma}{\partial t} = 
- \Omega_B \, \frac{\partial \Sigma}{\partial \varphi}
- \Sigma_{0B} \, f_1 \, \frac{\partial g_1}{\partial t}
- \Sigma_{0B} \, f_2 \, \frac{\partial g_2}{\partial t}  .
\end{equation}
Further on in the Tremaine-Weinberg method, the continuity equation is 
integrated over $x$ and $y$, which in the case considered here leads to
\begin{eqnarray}
\lefteqn{\Omega_B \int_{-\infty}^{+\infty} \Sigma(x,y) \; x \; dx -
\Sigma_{0B} \, \frac{\partial g_1}{\partial t} \int \int f_1\; dx \; dy -}\\
 & & \Sigma_{0B} \, \frac{\partial g_2}{\partial t} \int \int f_2\; dx \; dy
= \int_{-\infty}^{+\infty} \Sigma(x,y) \; v_y(x,y) \; dx , \nonumber
\end{eqnarray}
where $v_y$ is the velocity of the tracer along the $y$ axis. (A4) is the 
counterpart of (4) and (5), and since bar's pulsation is explicitly included 
here, it contains two additional correction terms (second and third term),
involving integration over $f_1$ and $f_2$. This integration,
although formally extending to infinity, in this case is limited to the regions
where bar's surface density is non-zero. Thus since $f_1$ and $f_2$ contain 
only rotated Cartesian coordinates, the integrals that involve them are finite,
and will be denoted as $I_{1B}$ and $I_{2B}$. Moreover, let us represent the 
pulsation of the bar by a simple oscillatory form of $g_1$ and $g_2$:
\begin{equation}
g_{1,2} = 1 \pm \epsilon_B \sin \omega t,
\end{equation}
where $\omega=2\pi/P$, and $\epsilon_B$ controls the amplitude of the 
oscillation. Then, with the use of notation from Section 2, (A4) 
takes the form
\begin{equation}
\Omega_B L_B X_B - \epsilon_B \omega \, \cos \omega t \; \Sigma_{0B}(I_{1B}-I_{2B}) = L_B V_B / \sin i
\end{equation}
Similar exercise can be done for the second bar, indexed by $S$, leading to 
an equation being a counterpart to (A6)
\begin{equation}
\Omega_S L_S X_S - \epsilon_S \omega \, \cos \omega t \; \Sigma_{0S}(I_{1S}-I_{2S}) = L_S V_S / \sin i
\end{equation}
The sum of these two equations gives (6), the equation of the extended 
Tremaine-Weinberg method, but here with correction terms, one for each bar.
In both (A6) and (A7), the correction term is the second term.

For each bar, the integrals $I_1$ and $I_2$ can be evaluated explicitly, 
indicating that $\Sigma_{0B,S} (I_{1B,S} - I_{2B,S})$ is about twice smaller 
than $X_{B,S}L_{B,S}$ throughout each bar. 
Numerical simulations indicate that the overall pulsation of the inner bar
is more noticeable than that of the outer bar, hence the magnitude of the 
coefficient $\epsilon$ should be larger for the inner bar. In the numerical
N-body simulations by Rautiainen et al.(2002), the axial ratio of the inner
 bar, $b/a$, varies roughly between 0.52 and 0.72. In the notation adopted 
here, this corresponds to $\epsilon_S=0.16$. Finally, 
$\omega \equiv 2(\Omega_S-\Omega_B)$ is always smaller than $2\Omega_S$.
Thus the magnitude of the second, correction term in (A7) should not exceed 
about 15\% of the magnitude of the first, leading term in the case of the
inner bar. For the outer bar, the ratio of $\omega/\Omega_B$ is larger, but
since the pulsation of that bar has smaller amplitude, $\epsilon_B$ is 
significantly smaller.

This example shows that periodic changes in time of surface density of the 
tracer of each bar in this bar's reference frame can be accommodated as
correction terms in the extended Tremaine-Weinberg formalism proposed in
this paper, given that the amplitude of these changes is sufficiently small.
This amplitude, as observed in numerical simulations, is indeed small enough,
and it leads to correction terms that are one order of magnitude smaller 
than the leading terms in the extended Tremaine-Weinberg equation (6).


\begin{thebibliography}{8.}
\addcontentsline{toc}{section}{References}
\bibitem{bv} Begelman M.C., Volonteri M., Rees M.J., 2006, MNRAS, 10.1111/j.1365-2966.2006.10467.x
\bibitem{cd} Corsini E.M., Debattista V.P., Aguerri J.A.L., 2003, ApJ, 599, L29
\bibitem{ee} Elmegreen D.M., Elmegreen B.G, Chromey F.R., Hasselbacher D.A., Bisssel B.A., 1996, AJ, 111, 2233
\bibitem{em} Emsellem E., et al., 2004, MNRAS, 352, 721
\bibitem{e0} Erwin P., 2004, A\&A, 415, 941
\bibitem{e1} Erwin P., Sparke, L.S., 2002, AJ, 124, 65
\bibitem{e2} Erwin P., Sparke, L.S., 2003, ApJS, 146, 299
\bibitem{f1} Friedli D., 1996, A\&A, 312, 761
\bibitem{h1} Hernandez O., Wozniak H., Carignan C., Amram P., Chemin L., Daigle O., 2005, ApJ, 632, 253
\bibitem{k1} Kuijken K., Fisher D., Merrifield M.R., 1996, MNRAS, 283, 543
\bibitem{l1} Laine S., Shlosman I., Knapen J.H., Peletier, R.F., 2002, ApJ, 567, 97
\bibitem{lg} Louis P.D., Gerhard O.E., 1988, MNRAS, 233, 337
\bibitem{m0} Maciejewski W., 2003, in: Contopoulos G., Voglis N. (eds.) Lecture Notes in Physics Vol. 626, Springer Verlag, Berlin, 91
\bibitem{m9} Maciejewski W., 2004, in: Block D.L., Puerari I., Freeman K.C., Groess R., Block E.K. (eds.) Astrophys. and Space Sci. Lib. Vol.319, Springer Verlag, Berlin, 175
\bibitem{ma} Maciejewski W., Athanassoula E., 2006, in preparation
\bibitem{m1} Maciejewski W., Sparke L.S., 1997, ApJ, 484, L117
\bibitem{m2} Maciejewski W., Sparke L.S., 2000, MNRAS, 313, 745
\bibitem{rs} Rautiainen P., Salo H., Laurikainen E., 2002, MNRAS, 337, 1233 
\bibitem{rw} Rand R.J., Wallin J.F., 2004, ApJ, 614, 142
\bibitem{sm} Sellwood J. A., Merritt, D., 1994, ApJ, 425, 530     
\bibitem{sw} Sellwood J. A., Wilkinson A., 1993, Rep. Prog. Phys., 56, 173
\bibitem{sh} Shlosman I., Frank J., Begelman M.C., 1989, Nature, 338, 45
\bibitem{sr} Sridhar S., 1989, MNRAS, 238, 1159
\bibitem{tw} Tremaine S., Weinberg M.D., 1984, ApJ, 282, L5     
\bibitem{w1} Wozniak H., Friedli D., Martinet L., Martin P., Bratschi P., 1995, A\&AS, 111, 115   
\end{thebibliography}
\end{document}